\documentclass{ifacconf}

\usepackage{natbib}
\usepackage{algorithmic}
\usepackage{graphicx}
\usepackage{textcomp}
\usepackage{xcolor}
\usepackage{bm}
\usepackage{array, threeparttable, booktabs} 

\newcolumntype{P}[1]{>{\centering\arraybackslash}p{#1}}
\usepackage{etoolbox}
\usepackage{multirow}
\usepackage{mathtools}
\usepackage{flexisym}
\usepackage{breqn}

\begin{document}
\begin{frontmatter}

\title{An Algorithmic Safety VEST For Li-ion Batteries During Fast Charging}

\thanks[footnoteinfo]{This work was funded by the National Science Foundation under Grant No. 1762247.}

\author[First]{Peyman Mohtat}
\author[First]{Sravan Pannala}
\author[First]{Valentin Sulzer}
\author[First]{Jason~B.~Siegel}
\author[First]{Anna~G.~Stefanopoulou}

\address[First]{University of Michigan, 
   Ann Arbor, MI 48105, USA \\(e-mail: \{{pmohtat\}}@umich.edu).}

\begin{abstract}
Fast charging of lithium-ion batteries is crucial to increase desirability for consumers and hence accelerate the adoption of electric vehicles. 
A major barrier to shorter charge times is the accelerated aging of the battery at higher charging rates, which can be driven by lithium plating, increased solid electrolyte interphase growth due to elevated temperatures, and particle cracking due to mechanical stress. Lithium plating depends on the overpotential of the negative electrode, and mechanical stress depends on the concentration gradient, both of which cannot be measured directly. Techniques based on physics-based models of the battery and optimal control algorithms have been developed to this end. While these methods show promise in reducing degradation, their optimization algorithms' complexity can limit their implementation. In this paper, we present a method based on the constant current constant voltage (CC-CV) charging scheme, called CC-CV$\eta \sigma$T (VEST). The new approach is simpler to implement and can be used with any model to impose varying levels of constraints on variables pertinent to degradation, such as plating potential and mechanical stress. We demonstrate the new CC-CV$\eta \sigma$T charging using an electrochemical model with mechanical and thermal effects included. Furthermore, we discuss how uncertainties can be accounted for by considering safety margins for the plating and stress constraints. 
\end{abstract}

\end{frontmatter}
\section{Introduction}

Reducing lithium-ion batteries' charging time is one of the main barriers to making electric vehicles more accessible and appealing to a wide range of the population. A full charge under five minutes can alleviate people's concerns with limited access to charging at-home or work. Furthermore, with the widespread availability of fast charging and algorithms to mitigate degradation, it is also possible to reduce the battery pack size and, in return, reduce the cost of electric vehicles.

Unfortunately, lithium-ion batteries degrade much faster (\cite{tomaszewska2019lithium}) at higher C-rates needed for fast charging. The accelerated aging is primarily caused by an increased amount of lithium plating and higher mechanical stress. Irreversible lithium plating consumes lithium and decreases the capacity. Moreover, a large amount of plated lithium in the form of dendrites is a serious safety concern for the battery as it can lead to internal short circuits and fires. High currents also lead to a large concentration gradient inside the particle, which results in a great amount of mechanical stress at the particle surface. Large stresses can increase particle cracking, loss of active material, and ultimately cause a reduction in capacity. Another important degradation mechanism contributing to capacity fade is the growth of the solid electrolyte interphase (SEI), which again consumes lithium inventory. This aging mechanism is largely unavoidable. However, it is widely accepted that temperature plays a significant role in the growth of the SEI. Therefore, preventing elevated temperatures can also limit the amount of SEI growth. Lastly, the degradation of the electrolyte at high voltages can also reduce the lithium-ion battery's performance and safety, which is mainly avoided by constraining the maximum voltage. At the same time, voltage limits define the cell capacity and are often prescribed by the cell manufacturer.

\begin{figure} [!b]
	\centering
	\includegraphics[width=\linewidth]{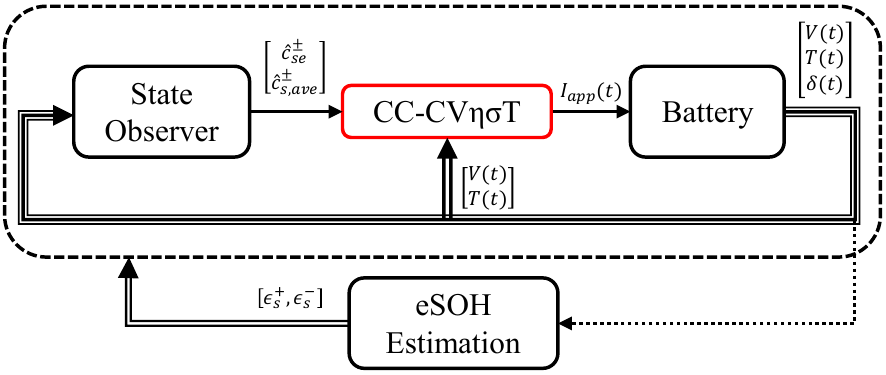}
	\caption{The proposed configuration for the CC-CV$\eta \sigma$T charging algorithm. The state observer and eSOH estimation methods are presented in earlier works (\cite{pannala2020improved,mohtat2019towards}).}
	\label{fig:sch}
\end{figure}

To reduce the degradation caused by lithium plating and stress, monitoring the plating potential and concentration distribution in the active material is needed. However, direct measurement of the variables mentioned above outside of a laboratory setting is essentially infeasible. Therefore, physics-based models are utilized instead. Many studies use models and optimization algorithms to minimize aging and charging time with a number of constraints. For example, in \cite{Klein2011} and \cite{perez2017optimal} the thermal and plating constraints are considered, and the charging problem is formulated using a nonlinear model predictive control scheme. This scheme is also utilized to minimize the aging due to SEI during fast charging in \cite{lam2020offline}. In \cite{Suthar2014} authors used simultaneous nonlinear programming approach with stress constraints on the electrode particle to develop a fast-charging algorithm. Nevertheless, the complexity of the optimization algorithm of these methods is a limiting factor for their implementation on board vehicles. The conventional way of charging batteries is to apply either a constant current (CC) or a constant power (CP) until the maximum allowable voltage is reached and then switch to a constant voltage (CV) phase until the current falls below a certain threshold. It can be shown that given a maximum current and a maximum voltage constraint, the CC-CV protocol is the fastest charging method in terms of minimizing only the charge time, for example by \cite{park2020optimal} for a Single Particle Model. Although the CC-CV protocol is easily implementable and used widely in practice, it is unaware of aging constraints discussed earlier. The CC-CV methodology has been used in \cite{patnaik2018closed} to develop a constant-temperature constant-voltage (CT-CV) charging technique, however, this method is unaware of the aging constraints.

This paper aims to leverage the existing CC-CV protocol and augment it with additional constraints that are cognizant of the degradation mechanisms. The new charging algorithm, derived in  Section~\ref{fast_charge}, can be used with any model to impose constraints on aging variables. The new constraints are added linearly to the integral current controller, which also includes a saturation limit and anti-windup scheme. In Section~\ref{model}, an electro-chemical-mechanical model is utilized to demonstrate the new charging algorithm by imposing constraints on plating potential, $\eta_{pl}$, the mechanical stress, $\sigma$, and the temperature, $T$. Thus, the new algorithm is named CC-CV$\eta \sigma$T (VEST), accordingly. Since the new algorithm relies on latent variables in a model, a state observer is also required. Fig.~\ref{fig:sch} shows an example of such configuration, and the results are presented in Section~\ref{results}. Furthermore, due to the aging number of parameters may need re-calibration from time to time. Previously, we have developed an observer in \cite{pannala2020improved}, and an electrode state of health (eSOH) estimation method in \cite{mohtat2019towards} utilizing the cell voltage and expansion that can be used in conjunction with the CC-CV$\eta \sigma$T charging as shown in Fig.~\ref{fig:sch}. Nevertheless, the focus of this paper is to showcase the CC-CV$\eta \sigma$T charging algorithm.


\section{Fast Charging Algorithm}
\label{fast_charge}
The proposed fast charging algorithm is presented in this section. The new algorithm is based on the constant current constant voltage (CC-CV) charging scheme. The CC-CV charging can be implemented in a number of ways. For example, a switch may be used to change from the CC phase to the CV phase once the voltage limit is reached. In this case, a reset might be needed for the integrator current controller once the switch happens. Another method for implementing the CC-CV charging is to use an integrator controller with output saturation and anti-windup:

\begin{equation}
\label{eq:CCCV1}
\begin{aligned}
I(t) = \int_0^t-[K_{I,V}e_V(\tau)&             \\+K_{aw}(I(\tau)&-\text{max}(I_{max},I(\tau)))]d\tau,
\end{aligned}
\end{equation}
where $I<0$ is charging, $V_d$ is the desired or maximum allowable voltage, the voltage error value: $e_V(t)=V_d-V_t(t)$, $I_{max}$ is the maximum allowable current, and $K_{I,V}$ and $K_{aw}$ are the gains for the voltage error and anti-windup, respectively. The setpoint or the initial condition is $I(0)=I_{max}$. The applied or output current is given by 
\begin{align}
\label{eq:CCCV2}
I_{app}=\text{max}(I,I_{max}).
\end{align}

The above implementation of the CC-CV charging will be used to construct the new charging algorithm with aging-aware constraints. Additional variables considered for the new charging algorithm are the plating potential, $\eta_{pl}$, the mechanical stress, $\sigma$, and the temperature $T$. The definitions of the plating potential and mechanical stress are given in Section \ref{model}. The temperature is assumed to be directly measurable from the surface of the lithium-ion battery. However, it is also possible to incorporate a thermal model of the battery and utilize the estimated temperature inside or at the battery center. The additional constraints are as follows:
\begin{subequations}
\label{eq:const}
\begin{align}
\eta_{pl}&>\eta_{pl,d},\\
\sigma &< \sigma_d,\\
T &< T_d,
\end{align}
\end{subequations}
where $\eta_{pl,d}$ is the desired minimum potential to reduce the lithium plating, $\sigma_d$ is the desired maximum stress to reduce the material fracture and failure, and $T_d$ is the desired maximum temperature to reduce aging due to the SEI growth. The charging protocol in (\ref{eq:CCCV1}) is modified by adding the error between the variable and the desired value in (\ref{eq:const}). The new charging algorithm, CC-CV$\eta \sigma$T, is as follows:
\begin{equation}
\begin{aligned}
\label{eq:cccvest}
I(t) = &\int_o^t-[K_{I,V}e(\tau)+\\
            &(V_t(\tau)<V_d)(\eta_{pl}(\tau)<\eta_{pl,d})K_{I,\eta}e_{\eta}(\tau)+\\
            &(V_t(\tau)<V_d)(\sigma_{d}<\sigma(\tau))K_{I,\sigma}e_{\sigma}(\tau)+\\
            &(V_t(\tau)<V_d)(T_d<T(\tau))K_{I,T}e_{T}(\tau)+\\
            &K_{aw}(I(\tau)-\text{max}(I_{{max}},I(\tau)))]d\tau+\\
            &(V_t(t)<V_d)(\sigma_{d}<\sigma(t))K_{P,\sigma}e_{\sigma}(t)+\\
            &(V_t(t)<V_d)(T_d<T(t))K_{P,T}e_{T}(t)
\end{aligned}
\end{equation}
where $K_{I,\eta}$, $K_{I,\sigma}$, and $K_{I,T}$ are the respective integral gains for the plating potential, mechanical stress, and temperature. The plating potential error is $e_{\eta}(t)=\eta_{pl}(t)-\eta_{pl,d}$. The mechanical stress error is $e_{\sigma}(t)=\sigma_{d}-\sigma(t)$. The temperature error is $e_{T}(t)=T_d-T(t)$. The $K_{P,\sigma}$ and $K_{P,T}$ are the proportional gains that are added to mitigate the output oscillations. Note that conditional terms are also considered for the additional variables so that they are only active if the voltage is below the maximum voltage. This ensures that overvoltage condition is avoided. Finally, the output current is given by (\ref{eq:CCCV2}), similarly to the CC-CV charging.

\section{The Aging Related Variables}
\label{model}
To demonstrate the new charging algorithm an electro-chemical-mechanical model developed in \cite{mohtat2020differential} is utilized. The summary of the model equations is presented in the appendix \ref{appendix}. In this section, the lithium plating potential and the mechanical stress are defined, and their connection to aging is discussed.

\subsection{Li Plating Model}
The Li plating potential is defined as the following:
\begin{flalign}
\label{eq:eta_pl}
    \eta_{pl}=\phi_s-\phi_e-V_f,
\end{flalign}
where $\phi_s$ denotes the solid potential and $\phi_e$ denotes the electrolyte potential. Here $V_f$ is the voltage drop across the SEI film.

A simple model of the plating reaction current (\cite{Yang2017}) is given by 
\begin{flalign}
\label{eq:j_pl}
j_{pl} = -\frac{i_{0,pl}}{F}\exp\left(\frac{-\alpha_{c,pl}F}{RT}\eta_{pl}\right),
\end{flalign}
where $i_{0,pl}$ is the exchange current density of Li deposition, and $\alpha_{c,pl}$ has a value of 0.5. An important outcome of (\ref{eq:j_pl}) is that to limit the amount of lithium plating the lithium plating potential, $\eta_{pl}$, needs to stay above zero. Therefore, the constraint for lithium plating can be written as
\begin{flalign}
\eta_{pl} > 0 ~\text{V}.
\end{flalign}

\subsection{Stress Model} \label{subsec:exp}

During cycling, the intercalation of lithium leads to changes in the lattice parameters. These changes lead to a volume change of a unit lattice volume, which is a function of the local lithium concentration on a macroscopic scale. The volume change is represented by a volumetric strain, which is denoted by $\Delta \mathcal{V}$. The detailed derivation of the expansion and stress is also presented in the Appendix. The hydrostatic stress $\sigma_h$ is given by:
\begin{equation}
\begin{aligned}
\label{eq:sigma_h}
    \sigma_h(r,t)=\frac{2E}{3(1-\nu)}\left(\frac{1}{R_p^3}\int_{0}^{R_p}\rho^2 \Delta \mathcal{V}(c_s(\rho,t))d\rho \right. \\\left. -\frac{\Delta \mathcal{V}(c_s(r,t))}{3} \right).
\end{aligned}
\end{equation}
The stress distribution across the particle radius is not uniform, making the process of defining a single constraint more complex. Furthermore, mechanical failure and fracture happens with repeated cycling and more often at the particle's surface. The stress at the surface goes from compressive during lithiation to tensile during delithiation. Therefore, to limit the mechanical degradation, a constraint should be considered for the stress both during charging and discharging. Nevertheless, we have only considered a constraint on the absolute value of the stress at the surface during charging: 
\begin{flalign}
\lvert \sigma_{h}(R_p,t) \rvert < \sigma_{d,m},
\end{flalign}
where $\sigma_{d,m}$ is the maximum allowable stress that can be defined by the user to limit the aging.

\section{Results and Discussion}
\label{results}
In this section, the ability of the CC-CV$\eta \sigma$T algorithm to limit the aging-related variables is exhibited by utilizing the physics-based model in the appendix \ref{appendix} with the aging-related parameters presented in the previous section. The constraints selected for the CC-CV$\eta \sigma$T are as follows:
\begin{subequations}
\label{eq:const2}
\begin{align}
\eta_{pl}&>0~\text{V},\\
\lvert \sigma_{h}(R_p,t) \rvert &< 92~\text{MPa},\\
T &< 40~^o\text{C}.
\end{align}
\end{subequations}
The above values are primarily selected for demonstrating a charging scenario, in which all the constraints become activated. The maximum voltage is set to 4.2 V, and the maximum current is 8 C (40 A). The simulation is also done once using the CC-CV charging method to compare the results of the CC-CV$\eta \sigma$T. Finally, the approaches to include safety margins for the constraints are discussed by analyzing the lithium plating potential's sensitivity to the active material ratio since the reduction of active material of the graphite electrode during aging can significantly affect the plating potential. 

\subsection{Gain Selection} \label{subsec:results_gain}

\begin{figure} [!t]
	\centering
    \includegraphics[width=\columnwidth]{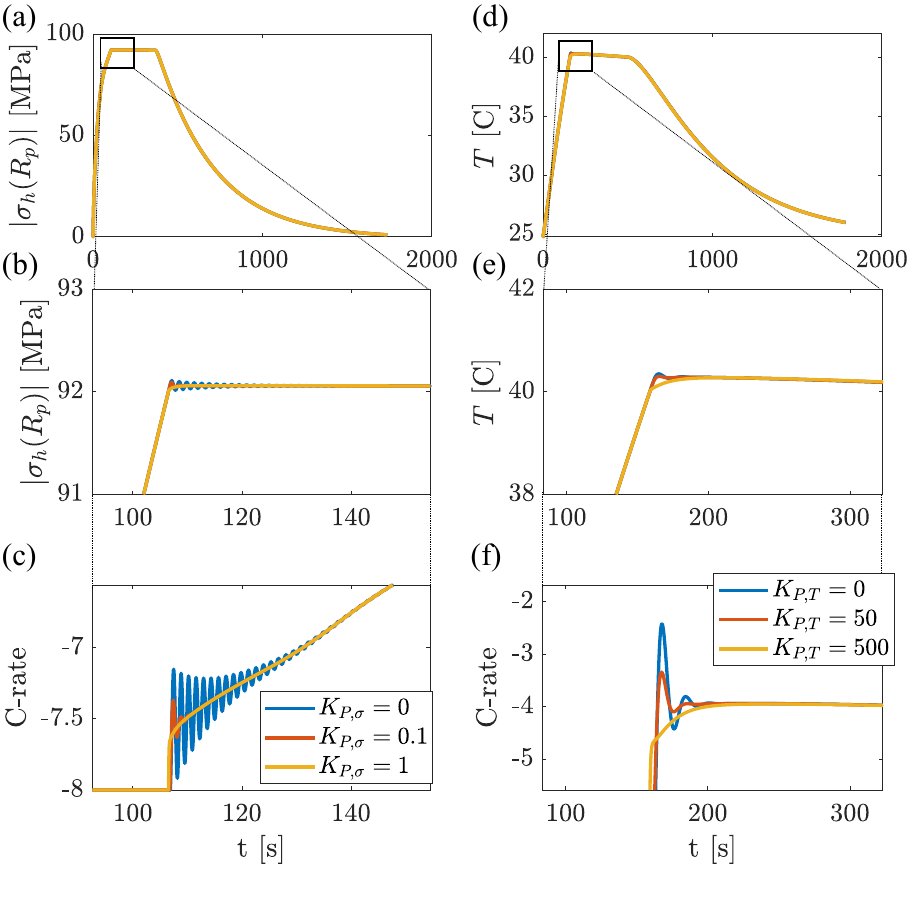}
	\caption{a) The mechanical stress, b) part of the response enlarged for clarity, and c) corresponding current signal for three different proportional gains. d) The temperature response, e) part of the response enlarged for clarity, and f) corresponding current for three different proportional gains. Note that the oscillations are reduced with the proportional controller.}
	\label{fig:gains}
\end{figure}

Determining the gains in (\ref{eq:cccvest}) requires careful consideration of the different operating points and the constraints. The desired response should have minimal oscillation and overshoot. However, since the lithium-ion battery is a highly non-linear system, guaranteeing the desired response with one set of gains is not feasible. Therefore, it is necessary to test the response at the different operating points and re-tune, if needed, in parameter drift or change during aging. Nevertheless, the gains shown in Table~\ref{tb:gains} are found to be highly effective under a wide range of operating points and are used in all the simulations unless specified otherwise.


\begin{table}[!b]
\begin{center}
\caption{Selected Gains}\label{tb:gains}
\begin{tabular}{c|P{0.65cm}P{0.65cm}P{0.65cm}P{0.65cm}P{0.65cm}P{0.65cm}P{0.65cm}}
Gain & $K_{I,V}$ & $K_{I,\eta}$ & $K_{I,\sigma}$ & $K_{P,\sigma}$ & $K_{I,T}$ & $K_{P,T}$ & $K_{aw}$\\\hline
Value &$50$ & $5e4$ & $200$ & $1$ &  $50$ & $500$ & $10$\\ 
Unit &$\frac{\text{A}}{\text{V.s}}$ & $\frac{\text{A}}{\text{V.s}}$ & $\frac{\text{A}}{\text{MPa.s}}$ & $\frac{\text{A}}{\text{MPa}}$ &  $\frac{\text{A}}{\text{K.s}}$ & $\frac{\text{A}}{\text{K}}$ & $\frac{1}{\text{s}}$\\ 
\end{tabular}
\end{center}
\end{table}

\begin{figure} [!t]
	\centering
    \includegraphics[width=\columnwidth]{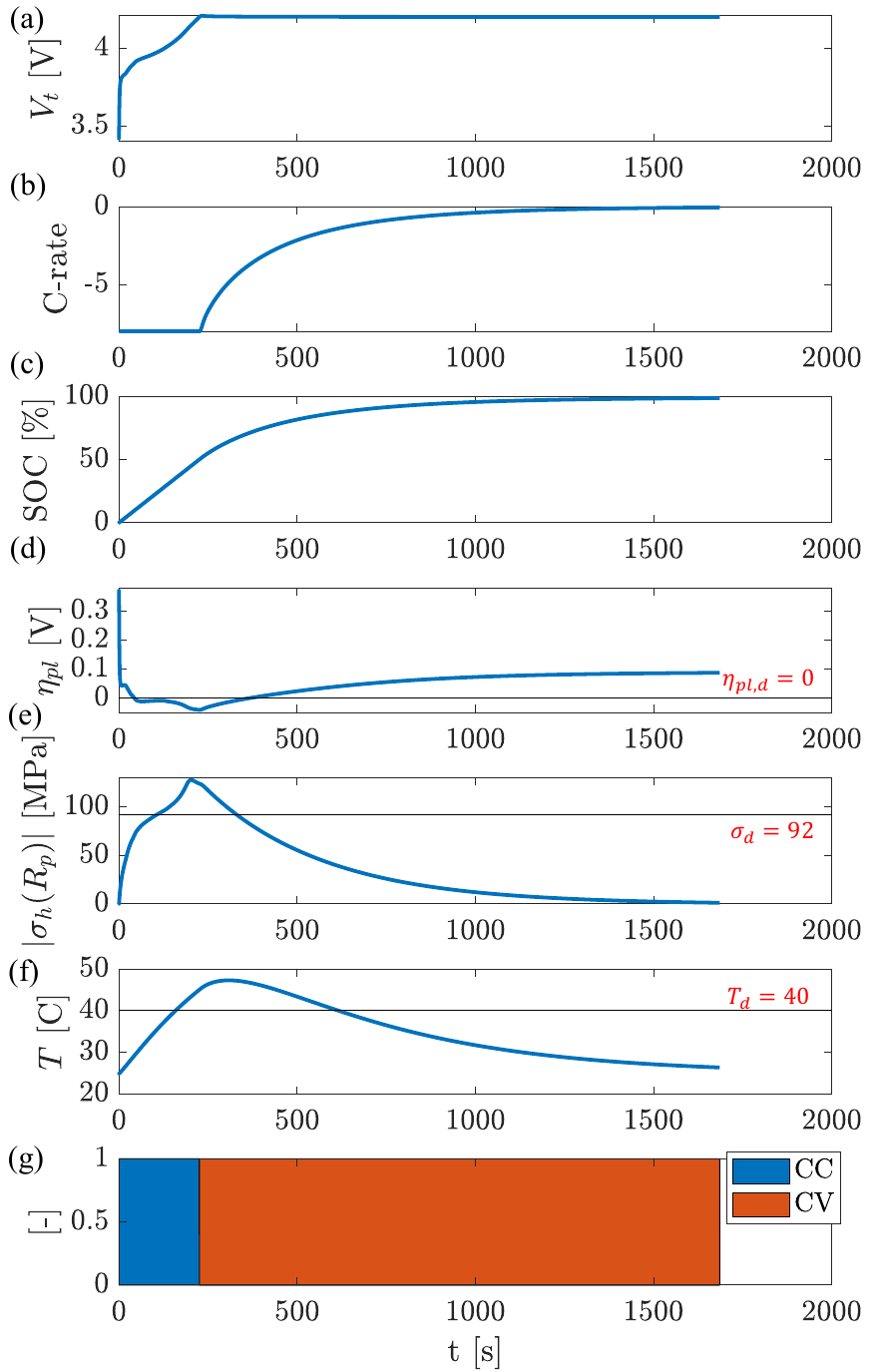}
	\caption{The CC-CV charging with 8 C and maximum voltage of 4.2 V. a) the voltage, b) the C-rate, c) the State of Charge (SOC), d) the plating potential, e) the stress, f) the temperature, and g) the charge-phase indicator. The limits shown in the plots for temperature, plating potential, and stress are violated.}
	\label{fig:cccv}
\end{figure}

Fig.~\ref{fig:gains} demonstrates the effects of the proportional controller and a possible tuning procedure. Two different scenarios are considered. In the first scenario depicted in Figs.~\ref{fig:gains} (a), (b), and (c) only the mechanical stress and voltage constraints are incorporated. The portion of the response highlighted in Figs.~\ref{fig:gains} (a) is shown in Figs.~\ref{fig:gains} (b) and the corresponding input current is shown in Figs.~\ref{fig:gains} (c). The simulation was performed with values of $K_{P,\sigma}=[0,0.1,1]$. The importance of including the proportional gain is shown, as the oscillations are entirely reduced by selecting an appropriate value for the $K_{P,\sigma}$ gain. 

In the second scenario, only the temperature and voltage constraints are considered. Similarly to the mechanical stress, the entire response for the temperature is shown in Figs.~\ref{fig:gains} (d) with the highlighted region shown in Figs.~\ref{fig:gains} (e) and in Figs.~\ref{fig:gains} (f) for the input current. The simulation was done with values of $K_{P,T}=[0,50,500]$. Again, as can be seen, the oscillations are reduced completely by adding the proportional gain.

\subsection{CC-CV} \label{subsec:results_cc}

The results of charging using the CC-CV method are shown in Fig.~\ref{fig:cccv}. As can be seen, the constraints considered in (\ref{eq:const2}) are violated as the CC-CV charging is not aware of these constraints. Therefore, repeated charging under this profile would lead to accelerated aging of the battery.

\subsection{CC-CV$\eta \sigma$T}

\begin{figure} [!t]
	\centering
    \includegraphics[width=\columnwidth]{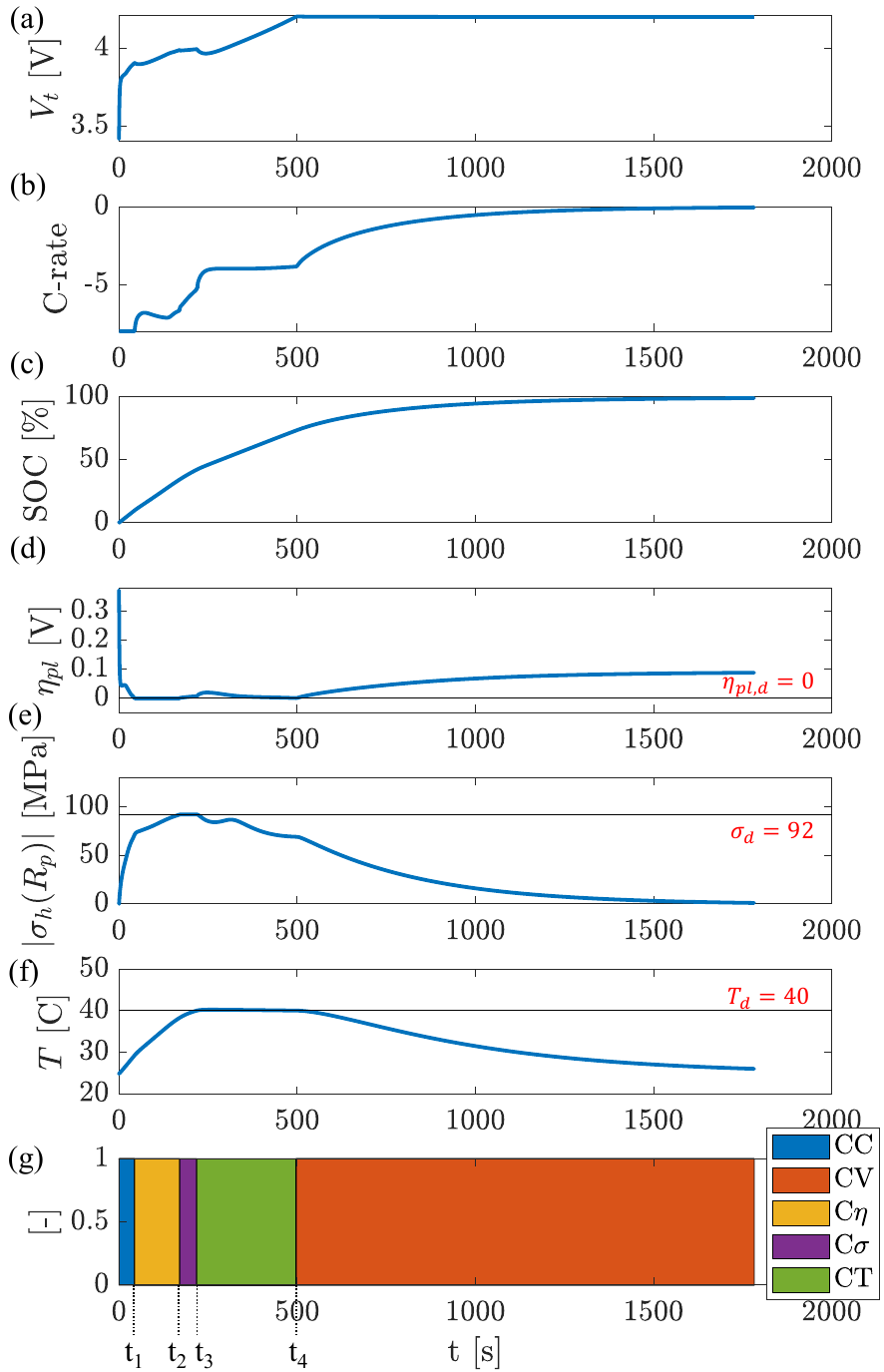}
	\caption{The CC-CV$\eta \sigma$T charging with 8 C, maximum voltage of 4.2 V, maximum temperature of 40$^oC$, minimum plating potential of 0 V, and maximum stress of 92 MPa. a) the voltage, b) the C-rate, c) the State of Charge (SOC), d) the plating potential, e) the stress, f) the temperature, and g) the charge-phase indicator.}
	\label{fig:cccvest}
\end{figure}

The results of charging using the CC-CV$\eta \sigma$T method and the constraints in (\ref{eq:const2}) are shown in Fig.~\ref{fig:cccvest}. As can be seen, the new algorithm maintained all the variables within the defined limits successfully. The charge-phase indicator shows when the constraints are active, and it is shown in Fig.~\ref{fig:cccvest} (g). At the time $t_1$, the plating potential reached the minimum, and the plating constraint was active. At the time $t_2$, the maximum stress was reached, and the stress constraint was active. At the time $t_3$, the maximum temperature constraint was activated. Finally, at time $t_4$, the maximum voltage was reached, and the CV phase was active.

The charge times using the CC-CV (Fig.~\ref{fig:cccv}) and {CC-CV$\eta \sigma$T} (Fig.~\ref{fig:cccvest}) algorithms are shown in Table \ref{tb:charge_time}. The charging time has increased with the new charging algorithm. However, while the charging time is only increased by 100 seconds, it is expected that the aging consequences of fast charging are reduced significantly.

\begin{table}[!h]
\begin{center}
\caption{Charge time}\label{tb:charge_time}
\begin{tabular}{ccc}
 SOC range & CC-CV & CC-CV$\eta \sigma$T \\\hline
0-80\% & 474 s & 580 s \\
0-100\% & 1685 s & 1783 s \\ \hline
\end{tabular}
\end{center}
\end{table}

\subsection{Sensitivity analysis for the plating potential}

\begin{figure} [!ht]
	\centering
    \includegraphics[width=0.95\columnwidth]{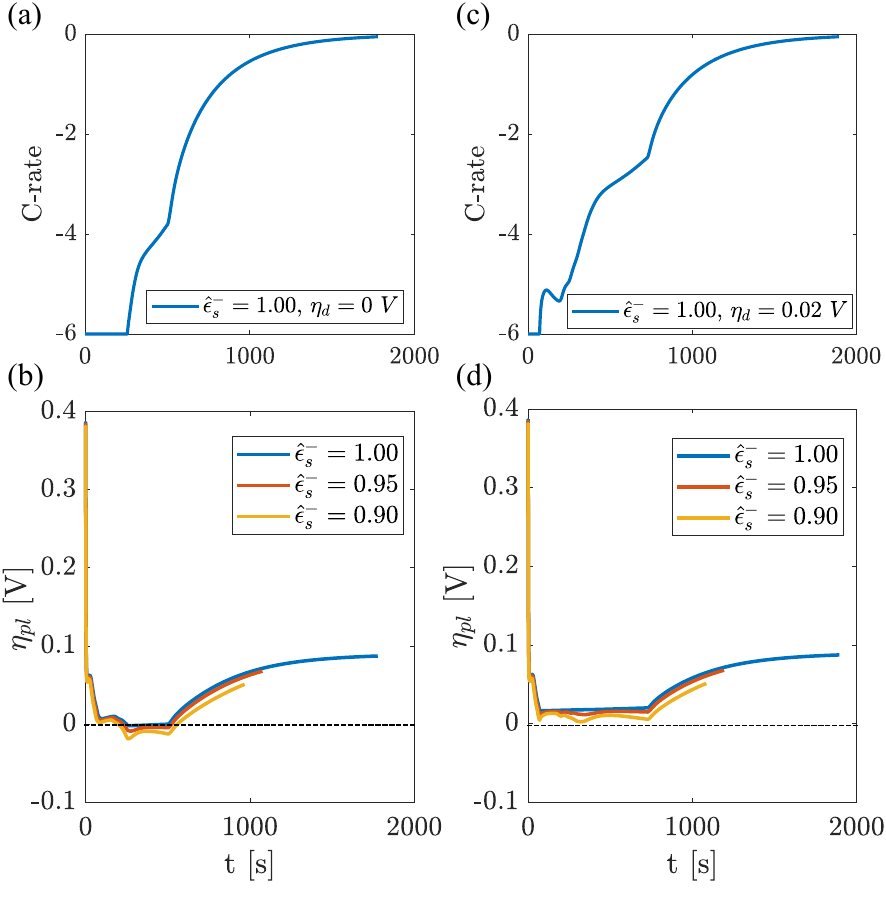}
	\caption{a) The input current based on fresh active material ratio cell (normalized active material ratio, $\hat{\epsilon}^-_s=1$), and 0 V plating potential constraint. b) The plating potential response assuming true active material ratios of 0.95 and 0.9. c) The input current based on the active material ratio of fresh cell, and a minimum plating potential of 0.02 V. d) The plating potential response with the normalized active material ratio of 0.95 and 0.9.}
	\label{fig:sense_plating}
\end{figure}

Errors in estimating the plating potential can stem from uncertainty in model parameters or the reduced-order model itself. Quantifying the effects of all these uncertainties is beyond the scope of this paper. Nevertheless, uncertainties can be accounted for by assuming a safety margin for the plating potential. The plating potential's sensitivity during the aging is demonstrated by assuming that the negative electrode's active material ratio, ${\epsilon}^-_s$, decreases during aging. The current input in Fig.~\ref{fig:sense_plating} (a) results from the new charging algorithm assuming the fresh active material ratio and a minimum plating potential of zero. However, as shown in Fig.~\ref{fig:sense_plating} (b), applying this current when the active material ratio is lowered results in the plating potential going below zero. This means that if the active material ratio is not re-calibrated, the drift in aging parameters can cause an error in the charging algorithm, which leads to more plating.

One way to account for uncertainties is to adjust the constraints. Fig.~\ref{fig:sense_plating} (c) shows the input current with a minimum plating potential of 0.02 V as a safety margin. As can be seen in Fig.~\ref{fig:sense_plating} (d), the plating potential remains above zero even when the active material ratio is reduce by 10\%. Therefore, the plating is prevented even at the aged conditions.

\section{Conclusion}
\label{conc}
In this paper, we have developed a new charging algorithm based on the CC-CV algorithm. The new algorithm can protect against aging related variables by imposing constraints on them. The new algorithm was used to limit the plating potential, mechanical stress, and temperature during charging, and therefore was named CC-CV$\eta \sigma$T (VEST). The new algorithm was tested using an electrochemical model with added mechanical and thermal dynamics. The results showed that new algorithm can limit these parameters successfully. The CC-CV$\eta \sigma$T is straightforward for implementation since it is based on the CC-CV logic. Future work will address the uncertainty quantification and gain selection process in a more methodical way.




\appendix

\section{Model Summary}
\label{appendix}
In the following equations the symbol $(*)$ is a place holder for either positive $(+)$ or negative $(-)$ electrodes. The number of particle sizes is denoted by $N_*^R$. The particle $(i)$ represents the particles with size $R_{p,i}^*$ (\cite{mohtat2020differential}).


\begin{subequations}
\begin{align}
\label{diff2}
\frac{\partial c_{s,i}^*}{\partial t}(r,t)&=\frac{1}{r^2}\frac{1}{\partial r}\left[D_{s,i}^*r^2\frac{\partial c_{s,i}^*}{\partial r}(r,t)\right],\\
\frac{\partial c_{s,i}^*}{\partial r}(0,t)&=0,\\
D_{s,i}^*\frac{\partial c_{s,i}^*}{\partial r}(R_{p,i}^*,t)&=-j_i^*(t),
\end{align}
\end{subequations}

\begin{subequations}
\begin{align}
\label{bv1}
j_i^*(t) = \frac{i_{0,i}^*(t)}{F}\left(e^{\frac{\alpha_a F}{RT}\eta_i^*}-e^{\frac{-\alpha_c F}{RT}\eta_i^*}\right),
\end{align}
\begin{gather}
\begin{multlined}
i_{0,i}^*(t) = \frac{1}{l_*} \int_{0}^{l_*} k_{0}^*(c_e(x,t))^{\alpha_a}(c_{s,max,i}^*-c_{se,i}^*(t))^{\alpha_a}\times \\(c_{se,i}^*(t))^{\alpha_c} dx,
\end{multlined}
\end{gather}
\end{subequations}


\begin{subequations}
\label{ch1}
\begin{align}
I = \sum_{i=1}^{N_*^R}l_* a_{s,i}^* Fj_{i}^*,~~~a_i^* = \frac{3\zeta_i^* \epsilon_s^*}{R_{p,i}^*}
\end{align}
\end{subequations}
\begin{gather}
\begin{multlined}
\label{elecconc3}
\epsilon_e^{**} \frac{\partial c_{e}}{\partial t}(x,t)=\nabla.(\bar{D_e}(\epsilon_e^{**})^{\gamma} \nabla c_e(x,t))+\\ \frac{1-\bar{t}_+^0}{F}\times
\begin{cases}
\frac{I(t)}{l_{-}} & 0\le x<l_{-},\\
0 & l_{-}\le x\le l_{-}+l_{sep} ,\\
\frac{-I(t)}{l_{+}} & l_{-}+l_{sep}<x\le l_{c},\\
\end{cases}
\end{multlined}
\end{gather}
where the symbol $(**)$ can be $(+)$, $(-)$, and $(sep)$ for positive electrode, negative electrode, and separator regions.
\begin{gather}
\begin{multlined}
\Delta \Phi_e(t)= - \left( \frac{l_{-}}{3(\epsilon^{-}_e)^{\gamma}}+ \frac{l_{sep}}{(\epsilon^{sep}_e)^{\gamma}} +\frac{l_{+}}{3(\epsilon^{+}_e)^{\gamma}}\right)\frac{I(t)}{\bar{\kappa}}+\\ \frac{2RT}{Fc_{e,init}}\bar{tf}(\bar{c}_{e}^{+}-\bar{c}_{e}^{-}).
\label{electvolt2}
\end{multlined}
\end{gather}

The output voltage can be written for any index $i=1,...,N_{+}^R$ and $j=1,...,N_{-}^R$, therefore
\begin{gather}
\begin{multlined}
\label{finalvt} 
V(t)=\eta_i^{+}(t)+U_{+}(c_{se,i}^{+}(t))+V_{f,i}^{+}(t)-\eta_j^{-}(t)\\-U_{-}(c_{se,j}^{-}(t))-V_{f,j}^{-}(t)+\Delta \Phi_e(t).
\end{multlined}
\end{gather}

A lumped temperature model is considered for the energy balance equation, which is given by
\begin{gather}
\begin{multlined}
\label{energy2s}
\rho C_c l_c \frac{dT}{dt}(t)=-h(T(t)-T_{a}) -I(t)V(t)\\-\sum_{*=+\& -} \sum_{i=1}^{N_*^R}a_i^*Fj_i^*(t)l_*[U_*(c_{se,i}^*(t))+V_{f,i}^*(t)\\-T(t)\frac{\partial U_*}{\partial T}(c_{se,i}^*(t))],
\end{multlined}
\end{gather}

 The stress-strain relationship with intercalation expansion is given by
\begin{flalign}
\label{d23}
\epsilon_{r}=\frac{1}{E}[\sigma_{r}-2\nu \sigma_{t}]+\frac{\Delta \mathcal{V}\left(c_{s}\right)}{3},\\
\epsilon_{t}=\frac{1}{E}[(1-\nu)\sigma_{t}-\nu \sigma_{r}]+\frac{\Delta \mathcal{V}\left(c_{s}\right)}{3},
\label{d24}
\end{flalign}
\begin{flalign}
\label{eq:stress_eqb}
    \frac{d\sigma_r}{dt}+\frac{2(\sigma_r-\sigma_t)}{r}=0.
\end{flalign}
\begin{flalign}
\label{eq:strain_r}
    \epsilon_r=\frac{du}{dr}, \\
\label{eq:strain_t}
    \epsilon_t=\frac{u}{r}.
\end{flalign}
We use (\ref{d23})-(\ref{eq:strain_t}) to generate the displacement equation:
\begin{flalign}
\label{eq:displ}
    \frac{d^2u}{dr^2}+\frac{2}{r}\frac{du}{dr}-\frac{2u}{r}=\frac{1+\nu}{1-\nu}\frac{d}{dr}\left(\frac{\Delta \mathcal{V}(c_s(r))}{3}\right),
\end{flalign}
with the boundary conditions:
\begin{flalign}
\label{eq:bc_1}
    u(0,t)=0, \\
\label{eq:bc_2}
    \sigma_r(R_p,t)=0.
\end{flalign}
The solution for the displacement, $u$, is given by 
\begin{gather}
\begin{multlined}
    u(r)=\frac{1+\nu}{1-\nu}\frac{1}{3r^2}\left(\int_{0}^{r}\rho^2 \Delta \mathcal{V}(c_s(\rho,t)) d\rho \right)+\\ \frac{1-2\nu}{1-\nu}\frac{2r}{3R_p^3}\left(\int_{0}^{R_p}\rho^2 \Delta \mathcal{V}(c_s(\rho,t))d\rho \right).
\end{multlined}
\end{gather}
Then, the radial and tangential stress are given by:
\begin{equation}
\begin{aligned}
\label{eq:sigma_r}
\sigma_r(r,t)=\frac{2E}{3(1-\nu)}\left(\frac{1}{R_p^3}\int_{0}^{R_p}\rho^2 \Delta \mathcal{V}(c_s(\rho,t))d\rho - \right. \\ \left. \frac{1}{r^3}\int_{0}^{r}\rho^2 \Delta \mathcal{V}(c_s(\rho,t)) d\rho \right) ,
\end{aligned}
\end{equation}
\begin{equation}
\begin{aligned}
\label{eq:sigma_t}
    \sigma_t(r,t)=\frac{E}{3(1-\nu)}\left(\frac{2}{R_p^3}\int_{0}^{R_p}\rho^2 \Delta \mathcal{V}(c_s(\rho,t))d\rho \right.\\ \left. + \frac{1}{r^3}\int_{0}^{r}\rho^2 \Delta \mathcal{V}(c_s(\rho,t)) d\rho -\Delta \mathcal{V}(c_s(r,t))  \right).    
\end{aligned}
\end{equation}

The hydrostatic stress is equal to
\begin{equation}
\begin{aligned}
\sigma_h(r,t)=\frac{\sigma_r(r,t)+2\sigma_t(r,t)}{3}.
\end{aligned}
\end{equation}


\bibliography{refs}

\end{document}


The model is based on the multi-particle model with electrolyte and details can be found in Ref. \cite{mohtat2020differential}. In the following equations the symbol $(*)$ is a place holder for either positive $(p)$ or negative $(n)$ electrodes. The number of particle sizes is denoted by $N_*^R$. The particle $(i)$ represents the particles with size $R_{p,i}^*$.


\begin{subequations}
\begin{align}
\label{diff2}
\frac{\partial c_{s,i}^*}{\partial t}(r,t)&=\frac{1}{r^2}\frac{1}{\partial r}\left[D_i^*r^2\frac{\partial c_{s,i}^*}{\partial r}(r,t)\right],\\
\frac{\partial c_{s,i}^*}{\partial r}(0,t)&=0,\\
D_i^*\frac{\partial c_{s,i}^*}{\partial r}(R_{p,i}^*,t)&=-j_i^*(t).
\end{align}
\end{subequations}

\begin{subequations}
\begin{align}
\label{bv1}
j_i^*(t) = \frac{i_{0,i}^*(t)}{F}\left(e^{\frac{\alpha_a F}{RT}\eta_i^*}-e^{\frac{-\alpha_c F}{RT}\eta_i^*}\right),
\end{align}
\begin{gather}
\begin{multlined}
i_{0,i}^*(t) = \frac{1}{l_*} \int_{0}^{l_*} k_{0}^*(c_e(x,t))^{\alpha_a}(c_{s,max,i}^*-c_{se,i}^*(t))^{\alpha_a}\times \\(c_{se,i}^*(t))^{\alpha_c} dx, 
\end{multlined}
\end{gather}
\end{subequations}


\begin{flalign}
I = \sum_{i=1}^{N_*^R}l_* a_{s,i}^* Fj_{i}^*.
\label{ch1}
\end{flalign}
\begin{gather}
\begin{multlined}
\label{elecconc3}
\epsilon_2^{**} \frac{\partial c_{e}}{\partial t}(x,t)=\nabla.(\bar{D_e}(\epsilon_2^{**})^{\gamma} \nabla c_e(x,t))+\\ \frac{1-\bar{t}_+^0}{F}\times
\begin{cases}
\frac{I(t)}{l_n} & 0\le x<l_n,\\
0 & l_n\le x\le l_n+l_s ,\\
\frac{-I(t)}{l_p} & l_n+l_s<x\le l_{c},\\
\end{cases}
\end{multlined}
\end{gather}
\begin{gather}
\begin{multlined}
\Delta \Phi_2(t)= - \left( \frac{l_n}{3(\epsilon^n)^{\gamma}}+ \frac{l_s}{(\epsilon^s)^{\gamma}} +\frac{l_p}{3(\epsilon^p)^{\gamma}}\right)\frac{I(t)}{\bar{\kappa}}+\\ \frac{2RT}{Fc_{e,init}}\bar{tf}(\bar{c}_{e,p}-\bar{c}_{e,n}),
\label{electvolt2}
\end{multlined}
\end{gather}

The output voltage can be written for any index $i=1,...,N_p^R$ and $j=1,...,N_n^R$, therefore
\begin{gather}
\begin{multlined}
\label{finalvt} 
V(t)=\eta_i^p(t)+U_p(c_{se,i}^p(t))+V_{f,i}^p(t)-\eta_j^n(t)\\-U_n(c_{se,j}^n(t))-V_{f,j}^n(t)+\Delta \Phi_2(t).
\end{multlined}
\end{gather}

where $U$ is the half-cell open circuit potential, and $V_{R}(x,t)=R_{f}Fj(x,t)$ is the voltage drop due to the film resistance.

The initial concentrations of the electrodes are given by
\begin{flalign}
\label{eq:csp_init}
c_{s,0}^+=c_{s,max}^+(SOC_0\times(y_{100}-y_0)+y_0)\\
\label{eq:csn_init}
c_{s,0}^-=c_{s,max}^-(SOC_0\times(x_{100}-x_0)+x_0)
\end{flalign}
where $SOC_0$ is the initial state of charge, $c_{s,max}$ is the maximum particle concentration, $y_{100},\,y_0$ are the positive electrode stoichiometric windows and $x_{100},\,x_0$ are the negative electrode stoichiometric windows defined by the voltage limits and electrode physical dimensions \cite{mohtat2019towards}.




A lumped temperature model is considered for the energy balance equation, which is given by
\begin{gather}
\begin{multlined}
\label{energy2s}
\rho C_p l_c \frac{dT}{dt}(t)=-h(T(t)-T_{a}) -I(t)V(t)\\-\sum_{*=p\& n} \sum_{i=1}^{N_*^R}a_i^*Fj_i^*(t)l_*[U_*(c_{se,i}^*(t))+V_{f,i}^*(t)\\-T(t)\frac{\partial U_*}{\partial T}(c_{se,i}(t))].
\end{multlined}
\end{gather}
where $\rho$ is the volume average density, $C_p$ is the volume averaged specific heat, $h$ is the heat transfer coefficient for convection, and $T_a$ is the ambient temperature. 

 The stress-strain relationship with intercalation expansion is given by
\begin{flalign}
\epsilon_{ij}=\frac{1}{E}[(1+\nu)\sigma_{ij}-\nu \sigma_{kk}\delta_{ij}]+\frac{\Delta \mathcal{V}\left(c_{s}(r,t)\right)}{3}\delta_{ij},
\label{d2}
\end{flalign}
where $\delta_{ij}$ is the Kronecker delta. It is assumed that the every active material particle is sphere, so a spherical coordinate system is utilized. Furthermore, due to the symmetry $\sigma_{\theta \theta}=\sigma_{\phi \phi}$, the normal strains in \cref{d2} can be written as
\begin{flalign}
\label{d23}
\epsilon_{rr}=\frac{1}{E}[\sigma_{rr}-2\nu \sigma_{\theta \theta}]+\frac{\Delta \mathcal{V}\left(c_{s}(r,t)\right)}{3}\\
\epsilon_{\theta \theta}=\frac{1}{E}[(1-\nu)\sigma_{\theta \theta}-\nu \sigma_{rr}]+\frac{\Delta \mathcal{V}\left(c_{s}(r,t)\right)}{3}.
\label{d24}
\end{flalign}
Where $\epsilon_{r}$ and $\epsilon_{t}$ are the particle strains in radial and tangential directions respectively. Stresses are represented by $\sigma_r$ and $\sigma_t$. $E$ is the Young's Modulus, and $c_s(r)$ is the concentration distribution inside the particle. The Stress equilibrium in the radial direction is given by:
\begin{flalign}
\label{eq:stress_eqb}
    \frac{d\sigma_r}{dt}+\frac{2(\sigma_r-\sigma_t)}{r}=0
\end{flalign}
Strain-displacement relationship is given by:
\begin{flalign}
\label{eq:strain_r}
    \epsilon_r=\frac{du}{dr} \\
\label{eq:strain_t}
    \epsilon_t=\frac{u}{r}
\end{flalign}
Then, we use \Crefrange{d23}{eq:strain_t} to generate the displacement equation:
\begin{flalign}
\label{eq:displ}
    \frac{d^2u}{dr^2}+\frac{2}{r}\frac{du}{dr}-\frac{2u}{r}=\frac{1+\nu}{1-\nu}\frac{d}{dr}\left(\frac{\Delta \mathcal{V}(c_s(r))}{3}\right)
\end{flalign}
with the boundary conditions:
\begin{flalign}
\label{eq:bc_1}
    u(0,t)=0 \\
\label{eq:bc_2}
    \sigma_r(R_p,t)=0
\end{flalign}
Integrating the displacement equation using the boundary conditions gives us a solution for the displacement $u$, which is given by 
\begin{gather}
\begin{multlined}
    u(r)=\frac{1+\nu}{1-\nu}\frac{1}{3r^2}\left(\int_{0}^{r}\rho^2 \Delta \mathcal{V}(c_s(\rho,t)) d\rho \right)+\\ \frac{1-2\nu}{1-\nu}\frac{2r}{3R_p^3}\left(\int_{0}^{R_p}\rho^2 \Delta \mathcal{V}(c_s(\rho,t))d\rho \right)
\end{multlined}
\end{gather}
We substitute this solution back in \Crefrange{d23}{eq:strain_t} to get the stresses. The radial stress and tangential stress are given by:
\begin{equation}
\begin{aligned}
\label{eq:sigma_r}
\sigma_r(r,t)=\frac{2E}{3(1-\nu)}\left(\frac{1}{R_p^3}\int_{0}^{R_p}\rho^2 \Delta V(c_s(\rho,t))d\rho - \right. \\ \left. \frac{1}{r^3}\int_{0}^{r}\rho^2 \Delta V(c_s(\rho,t)) d\rho \right) 
\end{aligned}
\end{equation}
\begin{equation}
\begin{aligned}
\label{eq:sigma_t}
    \sigma_t(r,t)=\frac{E}{3(1-\nu)}\left(\frac{2}{R_p^3}\int_{0}^{R_p}\rho^2 \Delta V(c_s(\rho,t))d\rho \right.\\ \left. + \frac{1}{r^3}\int_{0}^{r}\rho^2 \Delta V(c_s(\rho,t)) d\rho -\Delta V(c_s(r,t))  \right)    
\end{aligned}
\end{equation}

The hydrostatic stress is given by 
\begin{equation}
\begin{aligned}
\sigma_h(r,t)=\frac{\sigma_r(r,t)+2\sigma_t(r,t)}{3}.
\end{aligned}
\end{equation}

\biblio